\documentclass[proceedings]{rmaa}
\usepackage{graphicx}
\usepackage[utf8x]{inputenc}
\usepackage{epstopdf}

\begin{document}

\title{Hardware and Software for a robotic network of telescopes - SONG}

\altaffiltext{1}{Stellar Astrophysics Centre (SAC), Aarhus University}
\altaffiltext{2}{Niels Bohr Institute, University of Copenhagen}
\altaffiltext{3}{Instituto de Astrof\'{i}sica de Canarias (IAC), Tenerife}

\author{M.~F.~Andersen,\altaffilmark{1}
F.~Grundahl,\altaffilmark{1}
J.~Christensen-Dalsgaard,\altaffilmark{1}
S.~Frandsen,\altaffilmark{1}
U.~G.~Jørgensen,\altaffilmark{2}
H.~Kjeldsen,\altaffilmark{1}
P.~Pall\'{e},\altaffilmark{3}
J.~Skottfelt,\altaffilmark{2}
A.~N.~Sørensen,\altaffilmark{2}
and
E.~Weiss,\altaffilmark{1}}

\resumen{ El proyecto SONG busca establecer una red de peque\~nos telescopios por todo el mundo para observar las estrellas de manera ininterrumpida durante d\'ias, semanas y hasta meses.
Ac\'a describimos los aspectos fundamentales para la construcci\'on de una red de este tipo y como operaremos cada observatorio como parte de la red entera.
Los observatorios SONG trabajar\'an con autonom\'ia y pueden ser controlados completamente de manera remota. Algunos elementos b\'asicos para su funcionamiento, como el hardware y software, ser\'an descritos. }
\abstract{
SONG aims at setting up a network of small 1m telescopes around the globe to observe stars uninterrupted throughout days, weeks and even months.
This paper describes the fundamental aspects for putting up such a network and how we will operate each site as part of the full network.
The SONG observatories will be working autonomously and automatic and can be fully controlled remotely.
}

\addkeyword{ Lucky Imaging }
\addkeyword{ Network }
\addkeyword{ Oscillations }
\addkeyword{ Spectrograph }
\addkeyword{ Stars }

\shortauthor{M. F. Andersen et. al.}
\shorttitle{Stellar Observations Network Group}

\maketitle

\section{Introduction}
Stellar Observations Network Group (SONG) is a Danish-lead initiative to construct a global network of small 1m telescopes around the globe. The primary idea is to place 4 identical telescopes in each hemisphere to be able to observe single objects continuously for days, weeks and even months. During night time stellar objects will be the primary targets and during day the Sun will be observed using a optical fibre pointed directly towards the Sun. \\ 
The scientific aims of the SONG project are to contribute to a deeper understanding of stellar structure and evolution by observing stellar oscillations. With high-precision radial-velocity measurements obtained with a high-resolution echelle spectrograph this will be possible for the brightest stars. Asteroseismology will be used to obtain the characteristics of stellar pulsations to infer the physical parameters such as age, mass and size of the observed stars. To be able to use asteroseismlogy long uninterrupted time series are needed. SONG provides a facility that is capable of producing time series as long as required. A second aim is to contribute to the understanding and distribution of small exo-planets. Lucky imaging will be used to detect planets by gravitational micro-lensing towards the galactic center. This method will allow detection of low-mass exo-planets in orbits far away from their parent star. The high-precision radial velocity measurements will be sensitive to planets of high mass and in close orbits and detection of these types of planets will be a secondary outcome of the asteroseismic observations.\\
The SONG network will be dedicated to operate as a single instrument to focus the effort on projects which have not been feasible before with single-site observatories and the dependency of allocated observing time on each telescope. \\


\section{Telescope and instruments}
There are two main instruments on each SONG site; the SONG spectrograph and two lucky imaging (LI) CCD cameras. The two LIs are mounted at one of the two Nasmyth foci on the telescope. The spectrograph is placed in the separate air-conditioned shipping container inside a temperature-stabilized box. The light from the telescope can either go to the LI cameras or enter the vacuum-pumped coud\'e path which leads the light to the spectrograph.

\subsection{The SONG telescope}
\begin{figure}[h]\centering
\caption{Schematic design of the SONG telescope and Nasmyth unit}  
\label{figure:1}
\includegraphics[scale=0.15]{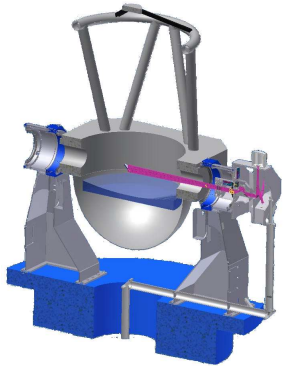}
\end{figure}

The SONG telescope is a 1m alt-az mounted telescope. The main mirror is 5cm thick and will be actively corrected for deformations based on Shack-Hartmann measurements. The secondary mirror is mounted in a movable hexapod for aligning and focusing and the third mirror can be rotated 180$^o$ to send the light to either of the two Nasmyth foci. The mount is capable of moving the telescope with speeds up to 20$^\circ$/s and a pointing model of less than 3\arcsec~RMS is reachable using around 30 stars. \\The dome is 5m in diameter and 7 side ports are placed around the dome pier to ensure good and fast ventilation of the dome. To minimize dome-seeing a cooling unit is installed to keep the temperature of the dome down during the day. On Figure~\ref{figure:1} a cut-through view of the telescope and part of the light path is shown.\\
With the combination of a 1m telescope and the SONG guiding system stars down to a V magnitude of around 7-8 will be possible to observe.

\subsection{The SONG spectrograph}
The SONG spectrograph is a high-resolution echelle spectrograph and normal operation resolution will be approximately 100.000 with a 1.2\arcsec~slit.\\
In Table~\ref{table:1} the basic parameters for the spectrograph are shown.\\
It is intended to be used in two main configurations:\\ 

\begin{table*}
\scriptsize
\caption{Specifications for the SONG spectrograph (Grundahl (2013))}         
\label{table:1}      
\begin{center}
\begin{tabular}{l l}        
\toprule                 
Resolution$^1$ & 35,000 - 115,000 \\
Wavelength coverage & 4400 - 6900  \AA \\
Pixel scale (average) & 0.02 \AA \\
Number of spectral orders$^2$ & 51 \\
Echelle grating & Newport, R4, 31.6l/mm, 100x340x50mm \\
Beam diameter and F-ratio & 75mm, F/6 \\
CCD detector & Andor Ikon-L, 2kx2k, with 2x20 pixels overscan, 13.5$\mu$m pixels \\
Readout speeds & (0.05, 1.0, 3.0, 5.0) Mega-pixels per second \\
\bottomrule                                   
\end{tabular}
\tiny \\$^1$ Average values over all orders. The resolution in the highest wavelengths can be up to 125.000.\\ $^2$ All orders are not fully covered on the CCD. The spectrograph was designed for a larger CCD.
\end{center}
\end{table*}

1. Together with a iodine reference cell. In this configuration the star light will pass through the iodine cell and iodine absorption lines will be superposed on top of the stellar spectrum.\\ 

2. Conventional Thorium-Argon observations, where a number of Th-Ar calibration spectre are acquired before and after a number of star exposures. \\\\
The spectrograph is placed inside an insulated temperature-regulated box which keeps the temperature stable to within 1/10 of a degree for long-term stability.
\begin{figure}[h]\centering
\caption{Schematic design of the SONG spectrograph}  
\label{figure:2}
\includegraphics[width=0.8\columnwidth]{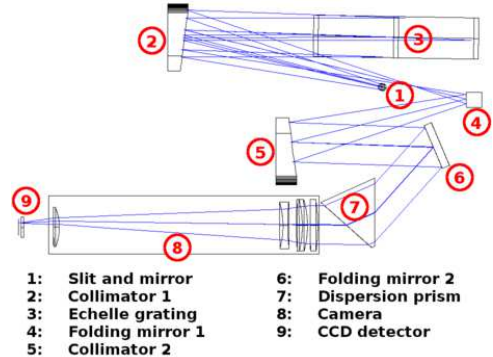}
\end{figure}
\\
A schematic view of the spectrograph can be seen in Figure~\ref{figure:2} where light enters at (1) and is collected on the CCD at (9).\\

\subsection{Lucky Imaging Cameras}
In one of the Nasmyth foci the two lucky imaging CCDs are mounted. An atmospheric dispersion corrector will automatically correct the incoming light for the wavelength-dependent dispersion of starlight through the atmosphere. An image de-rotator will eliminate image rotation caused by the alt-az mount. Two beam splitters will send the red part of the light to one CCD and the green part to the other. The blue part is send to a guide and focus monitoring camera. In Table~\ref{table:2} the specifications for the two LI CCDs are shown. 

\begin{figure}[h]\centering
\caption{Schematic view of the SONG Nasmyth unit box}  
\label{figure:4}
\begin{center}
\includegraphics[scale=0.35]{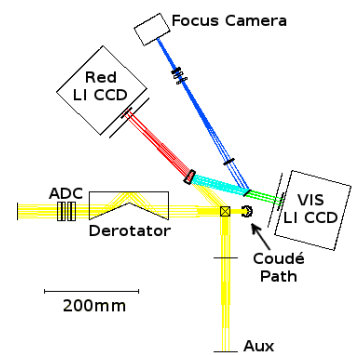}
\end{center}
\end{figure}

\begin{table}[h]
\scriptsize
\caption{Specifications for the SONG Lucky Imaging CCDs (Grundahl (2013))}         
\label{table:2}      
\begin{center}
\begin{tabular}{l l}
\toprule                
EMCCD Detector & Andor iXon model 897 \\
Active pixels & 512x512\\
Pixel size & 16x16$\mu$m \\
Max. read out speed & 33Hz (full frame) \\
Read out noise & $<$ 1 e$^-$ with EM gain\\
Field of view & $\sim$40\arcsec \\
Pixel scale & 0.08\arcsec per pixel \\
\bottomrule                                  
\end{tabular}
\end{center}
\end{table}


\section{Software}
The SONG observatories will operate fully automatically and without human interactions on site (except for maintenance and emergencies). To make the software development productive and efficient almost all devices are Ethernet controlled. The overall control of the instruments and devices will be carried out by a small number of software packages written in the programming language \textit{Python}.\\

\subsection{Database design}
\begin{figure}[h]\centering
\caption{SONG Database design}  
\label{figure:5}
\includegraphics[scale=0.30]{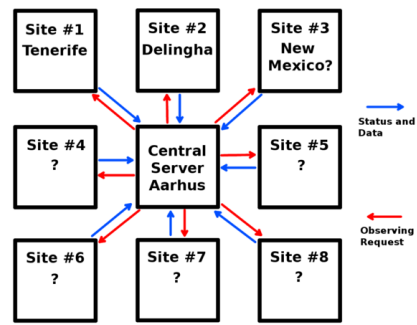}
\end{figure}

All SONG telescopes will primarily be operated as one single instrument when a network is completed. The idea is to have a number of database tables which are shared between every site. The SONG databases are created in PostgreSQL. A central server will keep tables on each site updated with new observing requests (OR) and the sites will return data and status of the observations carried out to the central server. See Figure~\ref{figure:5} for illustration. Specific database tables will be replicated to every site by the software package Slony-I$^1$\footnotetext[1]{http://slony.info}. Once observations have been carried out the status will be replicated and the data will be copied through GlusterFS$^2$\footnotetext[2]{http://www.gluster.org} to the central server. A number of other tables are copied from every site to the central server from where data can be displayed on dedicated web sites.\\  

\subsection{Observing procedure - Spectrograph}
The only human interaction when observing with SONG should be the decision of which object to observe, when and in what observing mode. This is combined into an OR which can be filled out through a web interface or by running a simple python script. The OR is then written into a database table on the central server from where the Slony mechanism will copy the OR to every site. On each site a local scheduler will pick up the newly inserted OR and execute it when certain conditions are fulfilled. When the scheduler decides to execute the OR it calls an observing script. From within this script the weather, position of object and time of day are checked and if all conditions are satisfied the telescope, all mirrors and filters will be moved to the specified values. When the telescope is tracking the object and mirrors are in place a spectrum will be acquired. When completed the status will be updated in a database table and checks of weather, object and time are performed again. This continues until some condition is no longer satisfied; object too low, weather turned bad, sunrise, OR finished etc.\\ Acquired data will be processed in real-time locally and in principle only the extracted velocities will be transferred to the central server. Raw data will be stored and backed up on site and only transferred to the central server if sufficient high speed network is available. One night of spectroscopic observations will produce of the order 10GB. Solar observation will produce about 300GB per day which will be stored on tape drives on site.\\


\section{Preliminary results}
\begin{figure}[h]\centering
\caption{Power spectrum of the Sun from 6 days of data}  
\label{figure:7}
\includegraphics[scale=0.40]{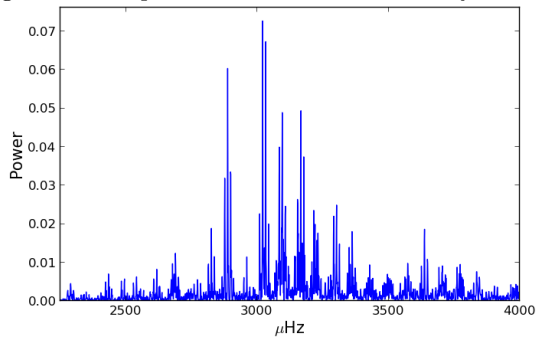}
\end{figure}

\begin{figure}[h]\centering
\caption{Power spectrum of the Sun from generated multisite data}  
\label{figure:8}
\begin{center}
\includegraphics[scale=0.40]{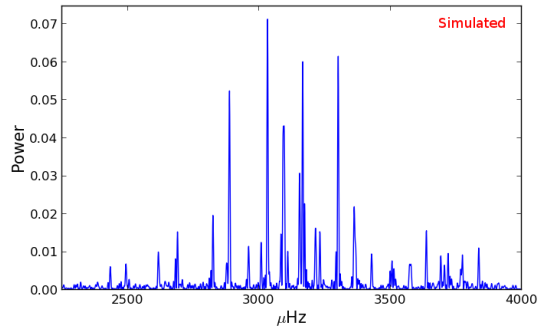}
\end{center}
\end{figure}

A 6 day campaign was carried out in June 2012 where an optical fibre was used to feed the SONG spectrograph with sunlight (Pall\'e (2013)). The fibre was mounted on a solar tracker and data were collected for about 10 hours on 6 consecutive days. On Figure~\ref{figure:7} the power spectrum of the 6 days of data is shown. The solar 5 minutes oscillations are clearly visible. Side peaks to each oscillation frequency caused by the gaps in the data are visible with a separation of 11.57$\mu$Hz (daily alias).
\\
For illustration purpose a 6 day data set was created without any gaps. The 40 most dominant frequencies in the above power spectrum were selected and used to create an artificial time series. White noise was added to the data set to make the two power spectra look alike in terms of noise. This data set illustrates the power of having a full network. On Figure~\ref{figure:8} the calculated power spectrum of the generated time series is shown. The side peaks corresponding to the daily gaps in the data are clearly gone and only the "real" frequencies are left.

\section{Status and future}
The SONG prototype telescope has been built and installed at Observatorio del Teide on the Canary island of Tenerife. The last tests of the telescope and instruments are being carried out at time of writing and full operation are expected to start early 2014.\\ 
A second node is under construction in China. The site is located near Delingha at the eastern edge of the Tibetan Plateau. The telescope has been manufactured, the observatory building has been built and the spectrograph had first light at the location of the assembly in September 2013. In early 2014 everything will be installed at the Delingha site and operation is expected to start in summer 2014.\\
At the time of writing only the two first sites are under development. Fund-raising for additional nodes is a constant ongoing process to extent the SONG network towards the goal of a full network. One additional node in North America would for specific targets complete the network in the northern hemisphere.

\section*{Acknowledgements}
We would like to acknowledge the Villum Foundation and the Carlsberg Foundation for the support on building the SONG prototype on Tenerife. The Stellar Astrophysics Centre is funded by The Danish National Research Foundation (Grant DNRF106) and research is supported by the ASTERISK project (ASTERoseismic Investigations with SONG and Kepler) funded by the European Research Council (Grant agreement n. 267864). We also gratefully acknowledge the support by the Spanish Ministry of Economy Competitiveness (MINECO) grant AYA2010-17803.

\end{document}